\documentclass[pra,twocolumn,showpacs]{revtex4}
\usepackage{graphicx}
\usepackage{dcolumn}
\begin{document}

\title{Parameterized optimized effective potential for atoms}

\author{A. Sarsa, F. J. G\'alvez and E. Buend\'{\i}a}
\affiliation{
Departamento de F\'{\i}sica Moderna, Facultad de Ciencias,
Universidad de Granada,  E-18071 Granada, Spain}

\date{\today}

\begin{abstract}

The optimized effective potential equations for atoms
have been solved by parameterizing the potential.
The expansion  is tailored
to fulfill the known asymptotic behavior of the effective potential at
both short and long distances. Both single configuration and multi 
configuration trial wave functions are implemented.
Applications to several atomic systems are
presented improving previous works. The results here obtained are very
close to those calculated in either the Hartree-Fock and the multi
configurational Hartree-Fock framework. 

\pacs{31.10.+z,31.15.-p,31.15.Pf,31.15.Ne,02.70.-c}

%31.10.+z Theory of electronic structure,
%         electronic transitions, and chemical binding
%31.15.-p Calculations and mathematical techniques in atomic and molecular
%         physics (excluding electron correlation calculations)
%31.15.Pf Variational techniques
%31.15.Ne Self-consistent-field methods
%02.70.-c Computational techniques

\end{abstract}

\maketitle

\newpage

\section{Introduction}
\label{sec.intro}

The Hartree-Fock (HF) method is the best approximation to the atomic
and molecular problem in the independent particle model with a single
configuration. Originally it was formulated
in terms of a single Slater determinant but in current applications a
single configuration, with one or more Slater determinants coupled in
the $LS$ scheme, is used, see e.g. \cite{ff97}. An alternative to the
HF equations, always within the independent particle model, is given
by the Optimized Effective Potential (OEP) method proposed by
\textcite{shho53} as a variational alternative to the Slater's
simplified treatment of the exchange term in the HF equations based on
the averaging in the occupied orbitals \cite{slater51}. 

In the OEP method an additional constraint is imposed into the variational
problem: the orbitals must satisfy a single-particle Schr\"odinger
equation with a certain local potential, the same for all the
electrons (with the same spin).  The expectation value of the
hamiltonian of the N-electron system becomes a functional of such a
local potential. The effective potential is then varied to minimize
the total energy. This gives rise to a linear
integral equation in the effective potential
\cite{shho53,tash76} whose solution gives the optimized effective
potential.  The wave function of the system is then
constructed from the single--particle wave functions which
are eigenfunctions of the optimized effective potential.  

The OEP method has been used in connection with the Kohn--Sham
density functional theory \cite{li91,engel92,hirata02,yawu02}. 
Accurate spin--polarized exchange--only Kohn--Shan potential has been 
constructed with the OEP method \cite{kli92a}. 
The local exchange potential obtained from
the OEP method has many of the analytical features of the exact Kohn-Sham
potential, and it has been recognized as the exact implementation of
exchange-only density functional theory \cite{gkkg00}. For example the
HF potential can support very few unoccupied excited states because of
its exponential fall off and therefore it is not a good starting point
to describe excited states. On the other hand as the OEP potentials presents
the proper long range behavior it provides better excitation energies.

The use and application of the original form of the OEP was hindered
because of the computational difficulties posed by the OEP integral
equation.  Initially, a numerical grid method was employed by
\textcite{tash76} to solve this equation and further more accurate
calculations for atoms have been carried out by using a more refined
mesh \cite{wang90,envo93}.  Also some approximations to the OEP
equations were proposed \cite{kli92b}. Very recently an iterative scheme based
on solving the partial differential equations satisfied by the orbitals shift
that satisfy exactly the  Krieger, Li and Iafrate approximation has been
devised \cite{kupe03}. An alternative methodology to
board this problem uses a finite basis set expansion  and/or 
analytical forms for the effective potential
\cite{yawu02,szgr74,dabe81,rwi88,gole94,fryu98,gorling99,ivanov99,cone01}.
The parameterization of the effective potential simplifies greatly the
numerical problem of solving the integral equations
\cite{yawu02,dabe81,fryu98,cone01}. The explicit form or the
parameterization is proposed by using different arguments
\cite{dabe81,fryu98,lasset85}. The
parameters can been fixed either by using the variational method or by
matching the one-body eigenvalues of the corresponding spin-averaged
Dirac equation to the experimentally observed one-electron ionization
energies \cite{rwi88}.

The aim of this work is to obtain the energy and other properties for
atoms in the parameterized OEP approximation. The analytical form 
used for the optimized potential has been used previously \cite{szgr74,rwi88} 
but with simpler expansions. In this work we increase the number of free
parameters in the effective potential until convergence is reached. In this
way we improve previous results obtained not only within the parameterized
OEP scheme but also by solving numerically the integral equation. We
have studied the ground state of the atoms Li to Ar in the $LS$ coupling.
Finally the OEP approximation is  extended here to multi configuration wave 
functions in the same spirit of the Multi Configuration Hartree Fock method.
The binding energy, single particle energies and exchange energies are obtained
and compared systematically with their Hartree-Fock counterparts.
The best results for the energy of the present work are roughly 1 mhartree
above the best self consistent field results showing the good performance
of the local potential approximation.

The structure of this work is as follows. 
In Section \ref{sec.method} we show in detail the parameterization and the 
algorithm used. 
The results obtained are reported and discussed in Section \ref{sec.results}. 
The conclusions and perspectives of this work can be found in 
Section \ref{sec.conclusions}. 
Atomic units are used throughout.

\section{Parameterized optimized effective potential}
\label{sec.method}

The single--particle wave functions used to construct the Slater
determinants of a given single configuration are  the eigenfunctions
of the so called effective potential, that in this work is taken to be
central,
\begin{equation}
\left(-\frac{1}{2}\vec{\nabla}^2 + V_{e}(r)\right) \, 
   \phi_{\lambda,\sigma}(\vec{r}) = 
\epsilon_{\lambda} \phi_{\lambda,\sigma}(\vec{r})
\label{schro}
\end{equation}
where $\lambda$ and $\sigma$ stand for the spatial and spin quantum
numbers, respectively. 

The total energy of the system is therefore a functional of this
single particle potential. The minimum condition on this potential leads to
the OEP equations \cite{shho53} fixing the best effective potential. The
energy so obtained will be an upper bound to the exact one and it will
be above the HF value. It is also worth pointing out 
that the role of the effective potential here is just an auxiliary function
used to calculate the orbitals in the wave function.

The parameterization chosen in this work for the  effective potential is 
\begin{equation}
V_e(r)=-\frac{1}{r} \left( Z-N+1+(N-1) \sum_{p=0}^S
\sum_{k=1}^{n_p} c_{k,p} r^p e^{-\beta_{k,p} r} \right)
\label{paref}
\end{equation}
with the condition
\begin{equation}
\sum_{k=1}^{n_0} c_{k,0}=1
\label{eq.constraint}
\end{equation}
imposed in order to match the correct short-range behavior of the
potential. This constraint makes that we must deal with, at least, two
terms with $p=0$, i.e. $n_0 \geq 2$. With respect to the long range
asymptotic behavior, this functional form is such that the potential
goes as $-(Z-N+1)/r$ for large electron-nucleus distances. The number
of basis functions used to expand the effective potential is
incremented systematically until convergence is reached.  As we shall
see later the importance of considering higher values of $p$ increases
with the number of electrons.  Finally it is worth mentioning here that
regular behavior close to the nucleus has shown to be relevant in
local effective potential theories \cite{pasa03}.
 
The algorithm is the following
\begin{enumerate}
\item Pick the initial values of the effective potential
parameters, $c_{k,p}$ and $\beta_{k,p}$ such that the short and
long range asymptotic behaviors are satisfied.
\item Solve the one-electron Schr\"odinger equation for each
occupied orbital.
\item Use these orbitals to build up the N-electron
wave function and calculate the expectation value of the
Hamiltonian.
\item Optimize the total energy with respect to the free parameters
($c_{k,p}$, $\beta_{k,p}$) with the constraint given by equation
\ref{eq.constraint}.
\item Increase the size of the expansion of the effective potential until
convergence.
\end{enumerate}
 
The single--particle Schr\"odinger equation \ref{schro} in step 3
is solved by expanding the radial orbitals in a basis set of Slater type 
functions as in the Roothaan-Hartree-Fock method. In particular we
have used the same size of the basis set as in reference \cite{clro74}: 
the $s$-type single particle orbitals are expanded by using 6 basis 
functions for the Li to Ne atoms and 8 basis  functions for the Na to 
Ar atoms. The radial part of $p$-type single particle orbitals is
developed as the sum of 4 basis functions for the B to Ne atoms,
5 basis functions for Na and Mg and 8 basis functions for the atoms
Al to Ar.  By doing this
all the different matrix elements can be computed analytically. The
total energy has been minimized with respect to the free parameters 
of the effective potential by using a SIMPLEX \cite{numrcip91}
algorithm.

Once obtained the single particle wave functions we can calculate some
other quantities of interest such as 
the exchange energy
\begin{equation}
E_{x} = -\frac{1}{2} \sum_{\sigma} \sum_{\lambda, \mu =1}^{N_{\sigma}}
\int {\rm d} \vec{r} \int {\rm d} \vec{r}^{\,\prime} \frac{\phi_{\lambda
\sigma}^\ast(\vec{r})\phi_{\mu \sigma}^\ast({\vec{r}^{\,\prime}}) \phi_{\mu
\sigma}(\vec{r})\phi_{\lambda \sigma}({\vec{r}}^{\,\prime})}{\mid\vec{r} -
\vec{r}^{\,\prime}\mid}
\end{equation}
and the Hartree--Fock single--particle energies, $\epsilon^{\rm
HF}_{nl}$, obtained from the expectation value 
\begin{equation}
\epsilon^{\rm HF}_{\lambda}
 = {\mathcal{I}}_{\lambda} + 
\sum_{ \mu}
(\mathcal{J}_{\lambda \mu} - {\mathcal{K}}_{\lambda \mu})
\label{singleener}
\end{equation}
where ${\mathcal{I}}, {\mathcal{J}}$ and ${\mathcal{K}}$ are the usual
single particle, direct and exchange terms calculated starting from
the eigenfunctions of the effective potential.  The single particle
energies $\epsilon^{\rm HF}_{\lambda}$ of equation \ref{singleener}
does not coincide with the eigenvalues $\epsilon_{\lambda}$ of
equation \ref{schro}, except for the highest occupied orbital. This
was proven in reference \cite{kli90} within a framework of
spin-polarized and arbitrary exchange-correlation functionals.  In the
scheme of the present study this condition states that $\epsilon^{\rm
HF}_{\lambda_h} = \epsilon_{\lambda_h}$, where $\lambda_h$ stands for
the highest occupied level.  The fulfillment of this 
condition has been used previously, e.g. \cite{li91,envo93} to asses 
the accuracy in the solution of the OEP equations.

In general the optimization procedure is very stable for both the
total energy and the Hartree--Fock values $\epsilon^{\rm
HF}_{\lambda}$. However this is not the case for the eigenvalue
$\epsilon_{\lambda}$, as has been noted previously
\cite{kli92a,envo93} by using a numerical grid method. 
For this reason, and because the exact solution of the OEP must
also satisfy the virial relation and  the exchange--only
virial relation  \cite{ghpa85,lepe85}
we have minimized the quantity 
\begin{equation}
\langle H \rangle + \mid 
                    \epsilon_{\lambda_h} - \epsilon^{\rm HF}_{\lambda_h}
		    \mid
+ \mid E_x-E_x^{vr}\mid
\end{equation}
where  $E_x^{vr}$ is given by 
\begin{equation}
E_x^{vr} = 
- \int {\rm d} \vec{r} \rho(\vec{r}) \vec{r} \cdot \vec{\nabla} 
    V_x(\vec{r}) 
\end{equation}
and $V_x$ is the exchange potential. 
By doing this no significant
changes are found for both the total energy and the Hartree--Fock
eigenvalues. One should expect that this procedure gives rise to a
better description of the asymptotic region of the optimized effective
potential. 
As it has been previously pointed out \cite{li91} these two conditions
are in some sense complementary. The main contributions to the 
quantities involved in the exchange only virial condition arise 
from the internal region of the atoms whereas the highest energy 
eigenvalue is governed by the outer region. By including these two 
conditions in the energy functional we observed a better and more 
stable convergence in the free parameters as compared to an
unconstrained minimization of $\langle H \rangle $. With the basis set 
used here both conditions are satisfied within one part in $10^{-6}$ hartree.

The method can be generalized straightforwardly to deal with a Multi
Configuration expansion. The starting trial wave function is written as a
linear combination of $m$ single configuration wave functions with the
total orbital angular momentum and spin of the state under study. The
hamiltonian is diagonalized in this set. The orbitals required to
build the different Slater determinants are obtained as the
eigenfunctions of the  single particle effective potential
containing several free parameters. The total energy
is minimized with respect to those parameters as before.  This more
general trial wave function will provide not only a better description
of the lowest energy state of a given symmetry but also a variational
approximation to the excited states because of the Hylleraas-Undheim
theorem which states that the eigenvalues constitute upper bounds to
the first $m$ bound states. These type of wave functions have been
recently used along with a correlation factor of Jastrow type to study some
excited states of the beryllium atom of and its isoelectronic series
\cite{gbs02,gbs03}. 
 
\section{Results}
\label{sec.results}

In Table \ref{table1} the results obtained by using different
parameterizations of the effective potential are analyzed. We show the
values for the Ne atom which is representative for the systems studied
here. We compare with the numerical optimized effective potential
results of Refs. \cite{envo93} and \cite{alt78} and with the
finite-basis set expansion of the effective potential of
Refs. \cite{fryu98,cone01}. The approximate values of reference
\cite{kli92a} are also reported. The HF results are taken from
\cite{bbb93} and are the benchmark values for the total energy
obtained from the OEP method.  We also report the exchange energy,
$E_{x}$, and the eigenvalues $\epsilon_{nl}$ and $\epsilon^{\rm
HF}_{nl}$.  The notation used for the parameterization is
$(0^{n_0},1^{n_1},2^{n_2},\ldots)$ where $n_p$ is the number of
functions of the type $e^{-\beta_{k,p}r} r^p/r$ used in the
expansion of the effective potential \ref{paref}.

\begin{table*}[ht]
\caption{\label{table1}
Convergence in the parameterization of the effective potential for the Ne atom.
The results are compared with the Hartree-Fock values (HF), two different
set of results obtained by solving numerically the optimized effective 
potential (NOEP) and with the Krieger, Li and Iafrate approximation.
$(0^{n_0},1^{n_1},2^{n_2},\ldots)$ stands for $n_p$ functions of the
type $e^{-\beta_{k,p}r} r^p/r$ in the expansion.
}
\begin{ruledtabular}
\begin{tabular}{crrrrrr}
                    & HF\cite{bbb93}   & NOEP\cite{alt78}  & NOEP\cite{envo93} 
                    & KLI\cite{kli92a} & OEP\cite{fryu98}  & OEP\cite{cone01}\\
\hline
$\langle H \rangle$ & -128.54710       & -128.5455        & -128.5455  
                    & -128.5448        & -128.5455        & -128.5456 \\
$E_{x}$             & -12.10835        & -                & -12.1050   
                    & -                & -12.1068         & -12.1055 \\
\hline
$\epsilon^{\rm HF}_{1s}$ & -32.77244   & -                & -       
                         & -           & -                & -        \\
$\epsilon_{\rm 1s}$      & -           & -30.8155         & -
                         & -           & -                & -30.8274 \\
\hline
$\epsilon^{\rm HF}_{2s}$ & -1.93039    & -                & -         
                         & -           & -                & - \\
$\epsilon_{\rm 2s}$      & -           & -1.71200         & -         
                         & -           & -                & -1.7196  \\
\hline
$\epsilon^{\rm HF}_{2p}$ & -0.85041    & -                & -
                         & -0.8494     & -0.84975         & -\\
$\epsilon_{\rm 2p}$      & -           & -0.84571         & -0.8507   
                         & -0.8494     & -0.85210         & -0.8506\\
\hline
 & $(0^2)$   & $(0^3)$   & $(0^2,1^1)$ &  $(0^2,1^1,2^1)$ & $(0^2,1^2,2^2)$ &\\ 
\hline
$\langle H \rangle$ 
& -128.54486 & -128.54641 & -128.54632 & -128.54642       & -128.54652 & \\
$E_{x}$             
& -12.09600  & -12.11006 & -12.10020   & -12.10002        & -12.10591 & \\
\hline
$\epsilon^{\rm HF}_{1s}$ 
& -32.76213  & -32.76843 & -32.77819   & -32.77775        & -32.77532 & \\
$\epsilon_{\rm 1s}$      
& -30.98356  & -30.77320 & -30.83773   & -30.90031        & -30.87242 &  \\
\hline
$\epsilon^{\rm HF}_{2s}$ 
& -1.93007   & -1.92862  & -1.93405    & -1.93386         & -1.93102 & \\
$\epsilon_{\rm 2s}$      
& -1.70548   & -1.71716  & -1.71928    & -1.71978         & -1.72224 & \\ 
\hline
$\epsilon^{\rm HF}_{2p}$ 
& -0.85251   & -0.84848  & -0.85384    & -0.85381         & -0.85130 &  \\ 
$\epsilon_{\rm 2p}$      
& -0.85251   & -0.84848  & -0.85384    & -0.85381         & -0.85130 & \\
\end{tabular}
\end{ruledtabular}
\end{table*}

The best results are obtained with the $(0^2,1^2,2^2)$
parameterization, and this is the one used for the rest of the
atoms. It can be seen that the value of the total energy does not
depend substantially on the basis set, but this is not the case of the
other quantities.  The use of bigger basis set
sizes for the effective potential does not improve noticeably the
energy   for the atoms Li to Ar and it increases the computational effort.
However for heavier atoms we have numerically checked that 
the rate of convergence can be substantially
improved by including higher powers of $r$ in the parameterization of the
effective potential.

The best results of this work improve previous ones obtained within
the optimized effective potential scheme and it is only 0.57 mhartree
above the Hartree-Fock result. The expectation values $\epsilon^{\rm
HF}_{\lambda}$ obtained with the OEP orbitals are in a very good
agreement with the corresponding single particle energies obtained
within the Hartree-Fock framework. This is because of the fact,
pointed out previously \cite{kli92a}, that the single particle wave
functions calculated from the OEP method are a good approximation to
the Hartree-Fock orbitals.

\begin{table*}[ht]
\caption{\label{table2}
Energy and single particle energies for He to Ne
atoms (OEP). 
The results are compared with the Hartree-Fock values (HF) obtained from
reference \cite{bbb93} and with the numerical solution of the optimized 
effective 
potential equation of \cite{alt78}, the exchange energies for the closed 
shell atoms have been taken from \cite{engel92}.
} 
\begin{ruledtabular}
\begin{tabular}{crrrrrrrr}
          & Li($^2$S)  & Be($^1$S)  & B($^2$P)   & C($^3$P)    
	  & N($^4$S)   & O($^3$P)   & F($^2$P)   & Ne($^1$S) \\
\hline
$E$ & & & & & & & &  \\
HF        &  -7.43273  & -14.57302  & -24.52906  &  -37.68862 
          & -54.40093  & -74.80940  & -99.40935  & -128.54710 \\
NOEP      &  -7.4324   & -14.5725   & -24.5278   &  -37.6865  
          & -54.3980   & -74.8075   & -99.4075   & -128.5455 \\
OEP       &  -7.43261  & -14.57291  & -24.52874  &  -37.68774 
          & -54.40029  & -74.80888  & -99.40875  & -128.54652 \\
\hline
$E_{x}$ & & & & & & & &  \\ 
HF        & -1.78119   & -2.66692   &  -3.74759  &  -5.04930  
          & -6.59707   & -8.18189   & -10.01105  & -12.10835 \\
NOEP      & -          & -2.666     & -          & -         
          & -          & -          & -          & -12.105  \\
OEP       & -1.78119   & -2.66627   &  -3.74661  &  -5.05059 
          & -6.59606   & -8.18039   & -10.01267  & -12.10591\\
\hline
$\epsilon^{\rm HF}_{1s}$ & & & & & & & &  \\
HF        &  -2.47774  &  -4.73267  &  -7.69534  & -11.32552 
          & -15.62906  & -20.66866  & -26.38276  & -32.77244 \\
OEP       &  -2.47783  &  -4.73366  &  -7.69628  & -11.32370
          & -15.63409  & -20.66947  & -26.37617  & -32.77532 \\
$\epsilon_{1s}$ & & & & & & & &  \\
NOEP      &  -2.08186  &  -4.12785  &  -6.91330  & -10.35324 
          & -14.46742  & -19.21792  & -24.66731  & -30.81549 \\
OEP       &  -2.09495  &  -4.06912  &  -6.92455  & -10.42413 
          & -14.61828  & -19.30917  & -24.78314  & -30.87242 \\
\hline
$\epsilon^{\rm HF}_{2s}$ & & & & & & & &  \\
HF        & -0.19632   & -0.30927   & -0.49471   & -0.70563 
          & -0.94532   & -1.24432   & -1.57254   & -1.93039 \\
OEP       & -0.19630   & -0.30962   & -0.49556   & -0.70415 
          & -0.94748   & -1.24639   & -1.56989   & -1.93102 \\
$\epsilon_{2s}$ & & & & & & & &  \\
NOEP      & -0.19644   & -0.30838   & -0.52804   & -0.74958 
          & -0.99752   & -1.19129   & -1.43073   & -1.71200 \\
OEP       & -0.19630   & -0.30962   & -0.52968   & -0.75608 
          & -1.00990   & -1.20004   & -1.44285   & -1.72224 \\ 
\hline
$\epsilon^{\rm HF}_{2p}$ & & & & & & & &  \\
HF        & -          & -          & -0.30986   & -0.43334 
          & -0.56759   & -0.63191   & -0.73002   & -0.85041 \\
OEP       & -          & -          & -0.30844   & -0.43313 
          & -0.56816   & -0.63174   & -0.72885   & -0.85130 \\
$\epsilon_{2p}$ & & & & & & & &  \\
NOEP      & -          & -          & -0.30976   & -0.43067 
          & -0.56322   & -0.62933   & -0.72479   & -0.84571 \\
OEP       & -          & -          & -0.30844   & -0.43313 
          & -0.56816   & -0.63174   & -0.72885   & -0.85130 \\
\end{tabular}
\end{ruledtabular}
\end{table*}

\begin{table*}[ht]
\caption{\label{table3}
The same as in Table \ref{table2} for the atoms Na to Ar.
} 
\begin{ruledtabular}
\begin{tabular}{crrrrrrrr}
          & Na($^2$S)  & Mg($^1$S)  & Al($^2$P)  & Si($^3$P) 
	  & P($^4$S)   & S($^3$P)   & Cl($^2$P)  & Ar($^1$S)\\
\hline
$ E$  & & & & & & & &  \\
HF        & -161.85891 & -199.61464 & -241.87671 & -288.85436 
          & -340.71878 & -397.50490 & -459.48207 & -526.81751 \\
NOEP      & -161.8565  & -199.6115  & -241.873   & -288.850 
          & -340.714   & -397.500   & -459.477   & -526.810 \\
OEP       & -161.85770 & -199.61303 & -241.87415 & -288.85160 
          & -340.71537 & -397.50155 & -459.47854 & -526.81405 \\
\hline
$E_{x}$  & & & & & & & &  \\
HF        &  -14.01752 &  -15.99429 &  -18.07225 &  -20.28350  
          &  -22.64091 &  -25.00614 &  -27.51653 &  -30.18494 \\
NOEP      & -          &  -15.988   &  -         & -        
          & -          & -          &            &  -30.175  \\

OEP       & -14.01160  & -15.98762  & -18.06349  &  -20.27359  
          & -22.62640  & -24.99093  & -27.49826  &  -30.17936 \\
\hline
$\epsilon^{\rm HF}_{1s}$ & & & & & & & &  \\
HF        & -40.47850  & -49.03174  &  -58.50103 &  -68.81246 
          & -79.96971  & -92.00445  & -104.88442 & -118.61035 \\
OEP       & -40.47877  & -49.03776  &  -58.50569 &  -68.81693 
          & -79.97485  & -92.01615  & -104.89542 & -118.60723 \\
$\epsilon_{1s}$ & & & & & & & &  \\
NOEP      & -38.04248  & -46.32444  &  -55.5604  &  -65.6547 
          & -76.59185  & -88.34890  & -100.9622  & -144.4452 \\
OEP       & -37.91728  & -46.16018  & -55.37384  &  -65.48096 
          & -76.39372  & -88.09753  & -100.59527 & -114.45720 \\
\hline
$\epsilon^{\rm HF}_{2s}$ & & & & & & & &  \\
HF        & -2.79703   & -3.76772   & -4.91067   &  -6.15654 
          & -7.51110   & -9.00429   & -10.60748  & -12.32215 \\
OEP       & -2.79384   & -3.77025   & -4.91136   &  -6.15701 
          & -7.51443   & -9.01030   & -10.61285  & -12.31768 \\
$\epsilon_{2s}$ & & & & & & & &  \\ 
NOEP      & -2.48096   & -3.09619   & -4.15912   & -5.34036 
          & -6.63449   & -8.01012   & -9.51097   & -11.14504 \\
OEP       & -2.22265   & -3.11058   & -4.17483   & -5.355205 
          & -6.64244   & -7.97517   & -9.54963   & -11.19559 \\
\hline
$\epsilon^{\rm HF}_{2p}$ & & & & & & & &  \\
HF        & -1.51814   & -2.28223   & -3.21830   & -4.25605 
          & -5.40096   & -6.68251   & -8.07223   & -9.57146  \\
OEP       & -1.51617   & -2.28607   & -3.22178   & -4.25935 
          & -5.40745   & -6.69193   & -8.07954   & -9.56872 \\
$\epsilon_{2p}$ & & & & & & & &  \\
NOEP      & -1.19797   & -1.85896   & -2.73146   & -3.71956 
          & -4.81814   & -5.99544   & -7.29541   & -8.72568 \\
OEP       & -1.16901   & -1.86617   & -2.74466   & -3.73189 
          & -4.82825   & -5.96260   & -7.33117   & -8.76408 \\

\hline
$\epsilon^{\rm HF}_{3s}$ & & & & & & & &  \\
HF        & -0.18210   & -0.25305   & -0.39342   & -0.53984 
          & -0.69642   & -0.87953   & -1.07291   & -1.27735  \\
OEP       & -0.18172   & -0.25359   & -0.38663   & -0.54090 
          & -0.69896   & -0.88305   & -1.07617   & -1.27597 \\
$\epsilon_{3s}$ & & & & & & & &  \\
NOEP      & -0.18211   & -0.25206   & -0.38859   & -0.53937 
          & -0.69510   & -0.80812   & -0.93838   & -1.09280\\
OEP       & -0.18172   & -0.25359   & -0.38388   & -0.53904 
          & -0.69884   & -0.80924   & -0.94641   & -1.09886 \\
\hline
$\epsilon^{\rm HF}_{3p}$ & & & & & & & &  \\
HF        & -          & -          & -0.20995   & -0.29711 
          & -0.39171   & -0.43737   & -0.50640   & -0.59102  \\
OEP       & -          & -          & -0.21062   & -0.29721 
          & -0.39325   & -0.43950   & -0.50812   & -0.58979 \\
$ \epsilon_{3p}$ & & & & & & & &  \\
NOEP      & -          & -          & -0.20910   & -0.29630 
          & -0.38737   & -0.43662   & -0.50027   & -0.58532\\
OEP       & -          & -          & -0.21062   & -0.29721 
          & -0.39325   & -0.43950   & -0.50812   & -0.58979 \\
\end{tabular}
\end{ruledtabular}
\end{table*}

The ground state and exchange energies of the atoms Li to Ar as well
as the single--particle expectation values, $\epsilon^{\rm HF}_{nl}$,
and the single--particle OEP eigenvalues, $\epsilon_{nl}$, are
reported in tables \ref{table2} and \ref{table3}.
All of these quantities are compared
with those obtained within the HF framework and with the numerical
optimized effective potential results of reference \cite{alt78}.  Within the
Hartree-Fock framework the two quantities $\epsilon^{\rm HF}_{nl}$ and
$\epsilon_{nl}$ are the same so that we do not list them separately.
For all the atoms the $(0^2,1^2,2^2)$ parameterization for the
effective potential has been used.

\begin{figure}[ht]
\includegraphics[scale=0.65]{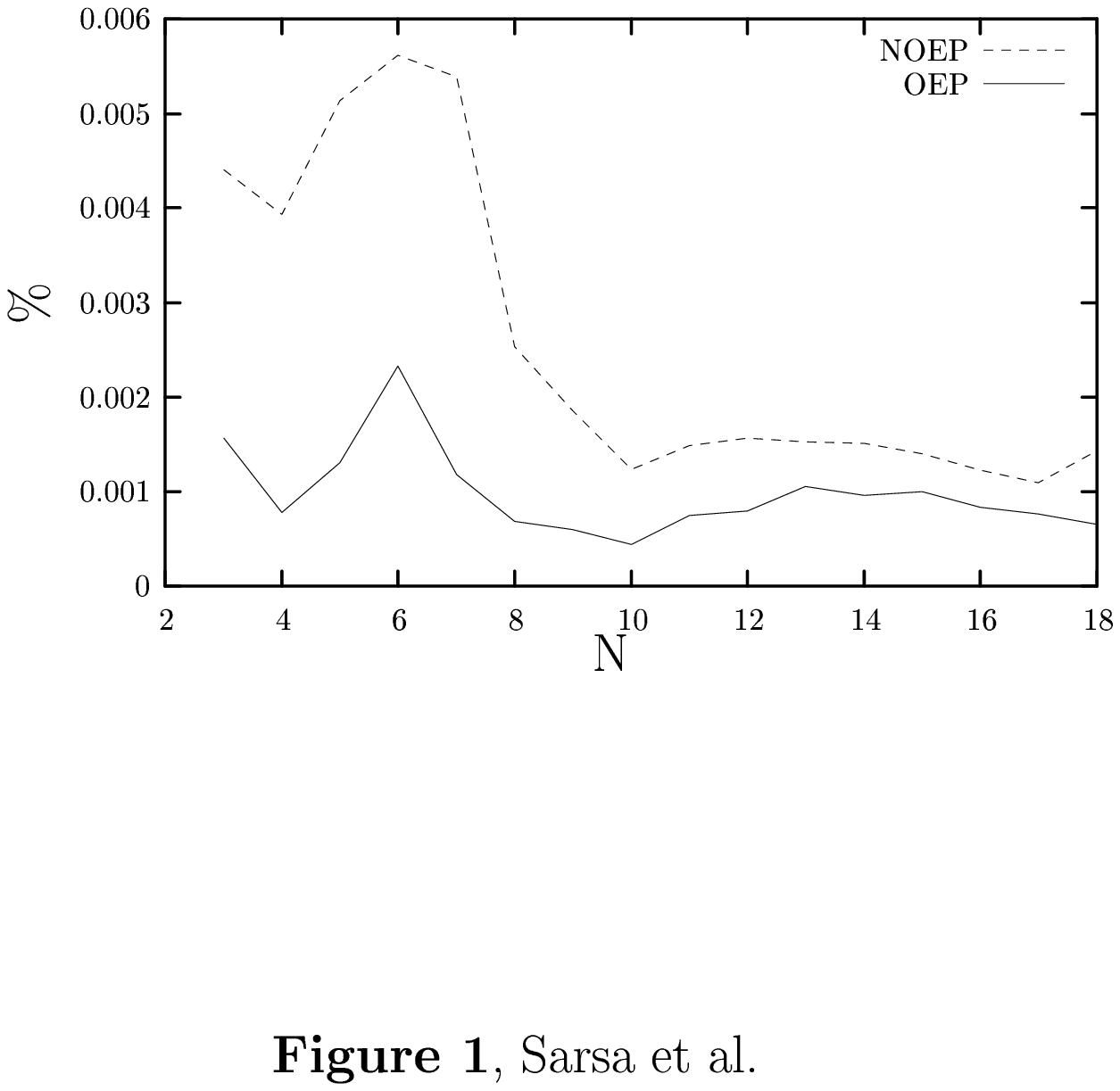}
\caption{\label{figure1} 
Relative difference (in percent) between the
OEP ground state energy of this work and the HF one for the atoms Li
through Ar as compared with the same quantity calculated from the
numerical results (NOEP) of \cite{alt78}.}
\end{figure}

As can be noted there is an appreciable improvement of the results,
especially for the lightest atoms, which can be more easily seen in
Figure \ref{figure1} where we plot the relative error (in $\%$) of our ground
state energy  with respect to the HF one as compared to the relative error 
for the numerical solution. In the present work the relative error is
nearly constant for all the atoms considered.  In principle one should
expect a better energy coming from the numerical solution than from the
parametrized one as the used in this work. 

Previous works have also reported energies calculated from a parametrized 
solution that are below the numerical solution \cite{fryu98,cone01}.
The reason for the better performance of the parametrized solution may 
be due to numerical inaccuracies in the numerical solution of the OEP 
equations due to the rather involved procedure of its solution. 
In an attempt to elucidate this fact we have carried out the following
calculation. Starting from the tabulation of \textcite{alt78} we have
built a parameterized potential of $(0^2,1^2,2^2)$ type by fitting the
numerical values to that functional form. Then the fitted potential
has been used in our code to determine the best orbitals within the effective
potential approach of this work without changing the fitted potential. Here
we show the results for the Ne atom, which are representative of the rest of
the atoms analyzed. The total energy obtained in this way is $-128.5457$ hartree
to be compared with $-128.5455$ hartree reported in
Ref. \cite{alt78}. The difference between these two values is due to a
better performance of the parameterized potential which allows one to
work analytically. In particular the asymptotic behavior of the
potential is taken into account in a more effective manner. 

\begin{figure}[ht]
\includegraphics[scale=0.65]{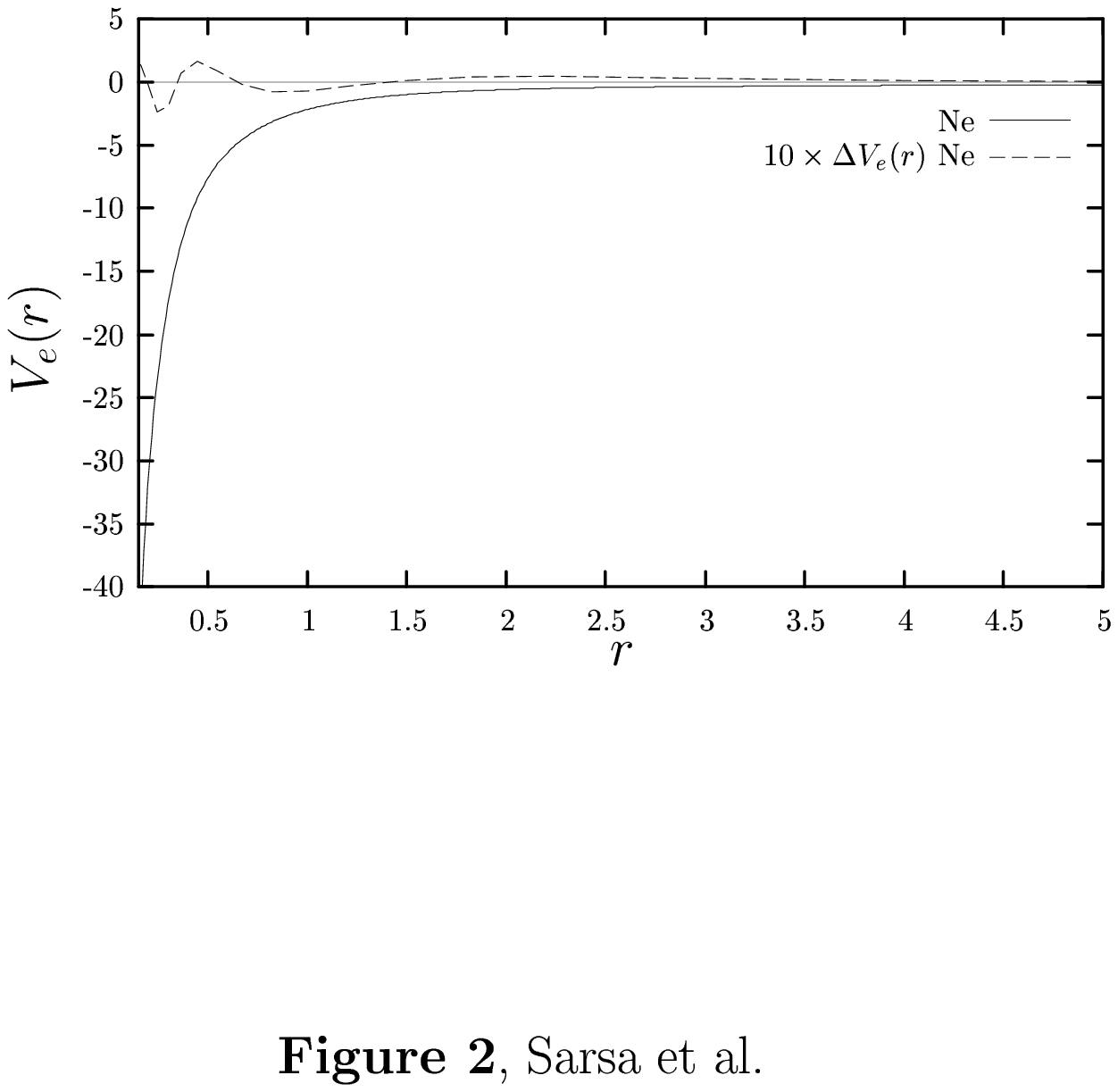}
\caption{\label{figure2} 
Effective potential for Ne  and the difference (multiplied by 10) with
the numerical results of \cite{alt78}.}
\end{figure}

The difference between the numerical OEP energy and the result
obtained here ($-128.5465$) is more accused. This discrepancy is due
to the differences in the effective potential. In Figure \ref{figure2}
we plot the effective potential for the Ne  atom and the
differences with the numerical results of \cite{alt78}. The results
for Ne are representative for the rest of the atoms considered
here. Notice that the differences are multiplied by 10 to better see
them in the scale of the figure.

\begin{figure}[ht]
\includegraphics[scale=0.65]{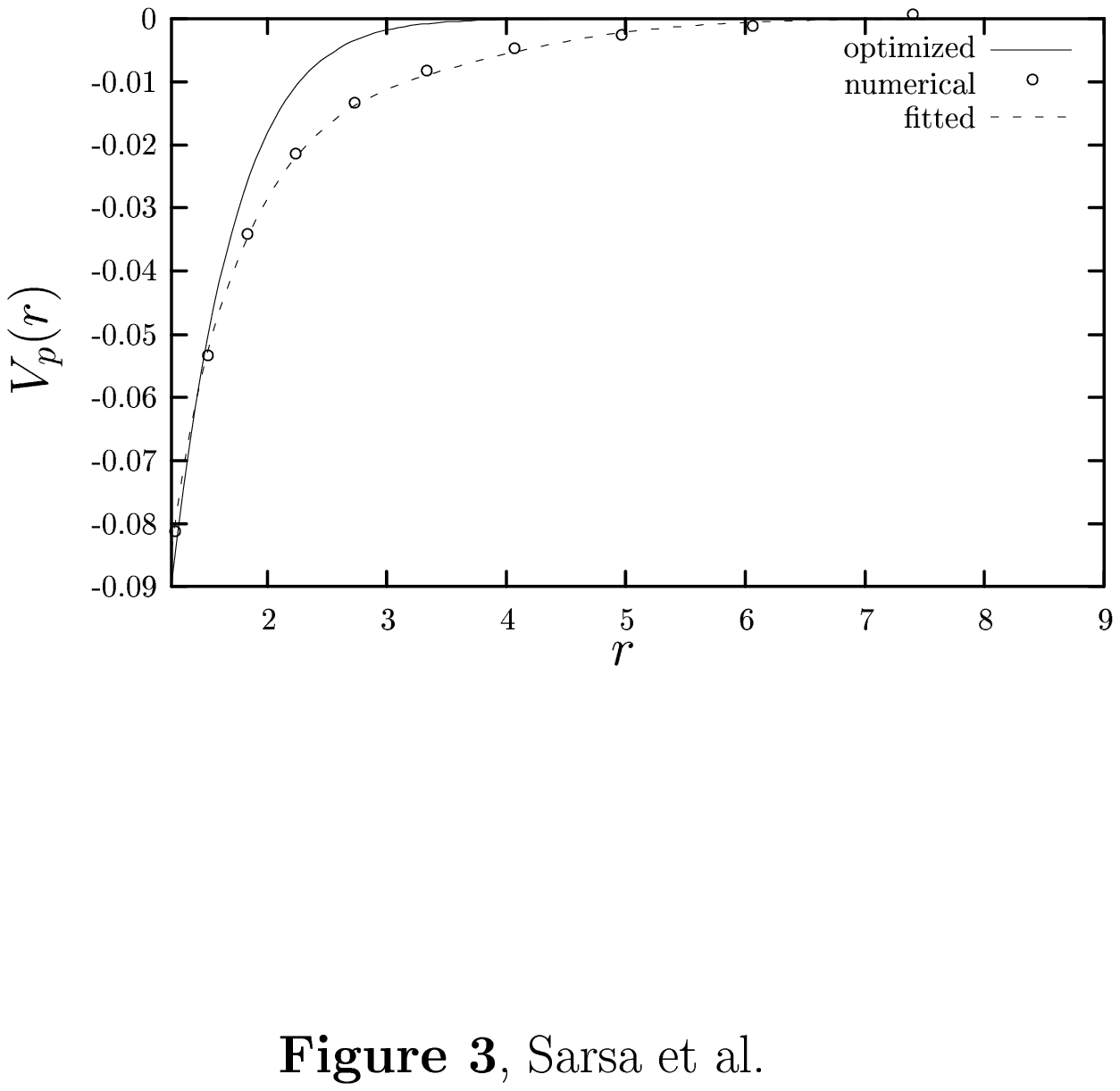}
\caption{\label{figure3} 
$V_p(r)$ defined in Eq. (\ref{vpot}) obtained from the best parameterized
effective potential of this work, {\em optimized}, and the numerical results
of Ref. \cite{alt78}, {\em numerical}. The analytical fit done to the numerical
values is also plotted, {\em fitted}.}
\end{figure}

In Figure \ref{figure3} we plot,
the function $V_p(r)$ obtained form the effective potential as (see
Eq. \ref{paref})

\begin{eqnarray}
\label{vpot}
V_p(r)=
\frac{r V_e(r)+Z-N+1} {(N-1)}
\end{eqnarray}

We compare the best optimized effective potential obtained in this work
with the numerical result calculated form the tabulated values of
Ref. \cite{alt78}, the fitted potential is included for the shake of
completeness.

Although the effective potential in the numerical approach and 
the parameterized solution of this work is very similar, see Figure 
\ref{figure1}, the differences for $V_p(r)$ become apparent. 
The main discrepancies are for $r \agt 1.5$ au, with a faster decay to 
zero of the parameterized solution.
The magnitude of this part of the effective potential is smaller
than the total potential. Therefore, the discrepancies between the
parametrized and numerical solutions are not very accused, in
this case $\sim 0.001$ mhartree.

It is worth mentioning here that for Be, Ne, Mg and Ar atoms more accurate 
numerical results for highest occupied single particle eigenvalues have 
been reported, \cite{li91,envo93}, but the total energy is very similar 
to that of reference \cite{alt78}.

To illustrate the performance of the Optimized Effective Potential 
with a multiconfigurational trial  wave function we have applied 
it to a simple case. In Table \ref{table4} we study the ground and 
the first $^1$P excited state of the 
beryllium atom using a two configuration wave function in both cases. We
compare our results with the corresponding Multi Configuration Hartree Fock
ones \cite{ff97,ff91} for the same states and configurations.
For the ground state, where the 2s-2p near degeneracy plays an
important role, we have expanded the wave function in terms of the
configurations 1s$^2$2s$^2$ and 1s$^2$2p$^2$. We have also studied the
$^1$P state arising from the 1s$^2$2s2p configuration by considering
the mixing with the 1s$^2$2p3d one.

Thus for the ground state, the trial wave function is written as
\begin{equation}
| \Psi \rangle  = c_1 | 1s^22s^2; ^1S \rangle  + c_2 | 1s^22p^2; ^1S \rangle
\end{equation}
and for the 1s$^2$2s2p $^1$P state the wave function is  
\begin{equation}
| \Psi \rangle  = d_1 | 1s^22s2p; ^1P \rangle  + d_2 | 1s^22p3d; ^1P \rangle
\end{equation}

In Table \ref{table4} we show the values for the energy and for the
coefficients $c_k$ and $d_k$ obtain
ed with the OEP method as compared
with the corresponding MCHF ones \cite{ff91}. 
It is apparent the good agreement between
the two sets of results for both states that illustrates that the OEP
method not only provides a good value for the energy but also an
adequate weight for any of the configurations involved in the
corresponding state. 
For this case the optimized effective potential method with 
multi configuration trial wave function compares with the MCHF 
at the same  level as the OEP method with the Hartree-Fock. 
More complex wave functions using a larger number of configurations 
have been used to study these and some other excited states of this 
atom and its isoelectronic series \cite{gbs02,gbs03}.
In those works, not fully 
optimized multi configuration wave function were used to build up 
more accurate explicitly correlated wave functions of Jastrow type.

\begin{table}[ht]
\caption{\label{table4}
Energy and coefficients for some multi configurational wave functions
for the ground and the first excited state of $^1$P type for the beryllium atom.
}
\begin{ruledtabular}
\begin{tabular}{crrr}
 $^1$S & E & $c_1$   & $c_2$  \\
\hline
MCHF  & -14.61685 & 0.95003  &  0.31214  \\
OEP   & -14.61637 & 0.95008  &  0.31202  \\
\hline 
 $^1$P & E & $d_1$  & $d_2$ \\
\hline
MCHF  & -14.41156 & 0.97524  & -0.22116 \\
OEP   & -14.41131 & 0.97489  & -0.22270  \\
\end{tabular}
\end{ruledtabular}
\end{table}

\section{Conclusions}
\label{sec.conclusions}

The optimized effective potential with parameterized potential has been
used to study the ground state of the atoms from Li to Ar. Parameterized
orbitals have been used to solve the corresponding single particle 
Schr\"odinger equation. 
The virial relation of Ghosh and Parr and Levy and Perdew 
involving the exchange energy and the exchange potential and 
a condition for the highest energy occupied orbital of Krieger, 
Li and Iafrate have been imposed. These are two analytically known 
conditions that the exact solution of the OEP equations must fulfill. 
We have included them in our functional and a constrained search of 
the optimum parameters is carried out. As a result for the minimum 
both of them hold within $10^{-6}$ hartree and the minimum energy 
is not substantially different to that obtained in an unconstrained
minimization. The tail of the effective potential is expected to 
be better reproduced by imposing the homo-condition.

The method has been generalized to work with
multi configuration wave functions as in the Multi Configuration Hartree
Fock method. This Multi Configuration OEP method provides
results very close to those obtained by using Multi Configuration
Hartree Fock (MCHF).

An analysis on the convergence on the parameterization of the effective
potential is carried out. The results obtained are very close to the
Hartree-Fock self consistent energies, eigenvalues and exchange energies
improving previous optimized effective potential calculations. Results for 
several bound states of the beryllium atom by using multi configuration 
wave functions have been reported.

\begin{acknowledgments}

This work has been partially supported by the Ministerio de Ciencia y
Tecnolog\'{\i}a and FEDER under contract  BFM2002-00200,  
and by the Junta de Andaluc\'{\i}a. 
 
\end{acknowledgments}

\end{document}